\documentclass[a4paper,11pt]{article}
\pdfoutput=1
\usepackage{jheppub} 
\usepackage[T1]{fontenc} 
\usepackage{amsmath,amssymb,graphicx,bbm,mathrsfs,nicefrac}
\usepackage{tikz}
\usepackage{amsmath}
\usepackage{amssymb}
\usepackage{graphicx,textcomp,float,gensymb,wrapfig, enumitem,comment,dsfont,framed,slashed,appendix,wrapfig,xcolor}
\usepackage{ bbold }
\usepackage{braket} 
\usepackage{ mathrsfs }
\usepackage[small]{caption}
\usepackage{subcaption}
\usepackage{etoolbox,hyperref,makecell}
\usepackage{feynmp}
\usetikzlibrary{arrows}
\usepackage{empheq}
\DeclareMathOperator{\Tr}{Tr}

\patchcmd{\abstract}{\null\vfil}{}{}{}
\newcommand{\bea}{\begin{eqnarray}}  
\newcommand{\eea}{\end{eqnarray}}

\title{Goldstone Inflation}
\author{Djuna Croon, Ver\'onica Sanz and Jack Setford}
\affiliation{Department of Physics and Astronomy, University of Sussex, Brighton BN1 9QH, UK}

\abstract{Identifying the inflaton with a pseudo-Goldstone boson explains the flatness of its potential. Successful Goldstone Inflation should also be robust against UV corrections, such as from quantum gravity: in the language of the effective field theory this implies that all scales are sub-Planckian.  
In this paper we present scenarios which realise both requirements by examining the structure of Goldstone potentials arising from Coleman-Weinberg contributions. We focus on single-field models, for which we notice that both bosonic and fermionic contributions are required and that spinorial fermion representations can generate the right potential shape. We then evaluate the constraints on non-Gaussianity from higher-derivative interactions, finding that axiomatic constraints on Goldstone boson scattering prevail over the current CMB measurements. The fit to CMB data can be connected to the UV completions for Goldstone Inflation, finding relations in the spectrum of new resonances. Finally, we show how hybrid inflation can be realised in the same context, where both the inflaton and the waterfall fields share a common origin as Goldstones.}
\emailAdd{d.croon@sussex.ac.uk}
\emailAdd{v.sanz@sussex.ac.uk}
\emailAdd{j.setford@sussex.ac.uk}

\begin{document}
\maketitle
\flushbottom

\section{Introduction}

The empirically well supported paradigm of cosmic inflation~\cite{inflation-idea} has a hierarchy problem from the perspective of particle physics. Parameterised in terms of a slowly rolling scalar field, the scale of inflation (from CMB data~\cite{Planck2015}) is exceeded by the field excursion (given by the Lyth bound~\cite{lyth}) by roughly two orders of magnitude:
\bea \Lambda^4 = \left(1.88 \times 10^{16} \text{ GeV} \right)^4 \left(\frac{r}{.10}\right) \,\,\,\,\text{and} \,\,\,\, \Delta \phi \geq M_p \sqrt{\frac{r}{4 \pi}}
\eea
where $r$ is the ratio of the tensor to the scalar power spectrum, and where $M_p = 2.435 \times 10^{18} \text{ GeV} $ is the reduced Planck mass.
Meeting both these conditions implies an exceptionally flat potential for the inflaton, which generically is radiatively unstable. 

Natural Inflation (NI)~\cite{here} offers a solution to this hierarchy problem by imposing a symmetry on the inflaton: the inflaton potential exhibits a shift symmetry  $\phi \rightarrow \phi + C$ with $C$ a constant, and therefore could protected from higher order corrections. The shift symmetry is realised by identifying the inflaton with the Goldstone boson (GB) $\phi$  of a broken global symmetry $G$ to its subgroup $H$ ($\phi \in G/H$). In turn, the GB obtains a potential through effects that render $G$ inexact. The resulting degree of freedom is therefore not an exact Goldstone boson, but a \emph{pseudo-Goldstone boson} (pGB). Different effects can lead to an inexact global symmetry; we reviewed the relevant mechanism in~\cite{DandVrock}.

The original and most popular NI model has an axion as the inflaton, the GB of spontaneously broken Peccei-Quinn symmetry~\cite{here}. The axion gets a potential through non-perturbative (instanton) effects. As shown in Ref.~\cite{witten} these effects lead to the characteristic $ \cos ( \phi/f )$ potential across models, where  $f$ is the scale at which $G$ is broken. To obtain the famous NI model one adds a cosmological constant term to impose the phenomenological constraint $V(\phi_{min}) = 0$, to obtain,
 \bea \label{vanilla} V(\phi) = \Lambda^4 (1 +  \cos \phi/f)\eea

Alas, the original NI model can only be successfully reconciled with the data from CMB missions for superplanckian scales of the decay constant: $ f =  \mathcal{O} ( 10 M_p)$. This is evidently a problem, because above the Planck scale one should expect a theory of Quantum Gravity (QG), and it is known that theories of QG in general do not conserve global symmetries~\cite{gravityandglobal}. Therefore one generically expects large contributions to the simple potential \eqref{vanilla}, as was shown recently in~\cite{transaxions}. Thus, one may conclude that vanilla NI is not a good effective theory.\footnote{It is found that it is only possible to maintain control over the backreaction in very specific configurations, such as~\cite{1409.1221}.}

Different proposals have been made to explain the super-Planckian decay constant while maintaining the simple potential \eqref{vanilla} and the explanatory power of the model. Among these are Extra-Natural inflation~\cite{extra-natural}, hybrid axion models~\cite{hybrid, Peloso}, N-flation~\cite{N-flation, Mazumdar}, axion monodromy~\cite{axion-monodromy} and other pseudo-natural inflation models in Supersymmetry~\cite{pseudonat}. 
These proposals usually focus on generating an effective decay constant $f_{\mathit{eff}}$ in terms of model parameters, such that $ f_{\mathit{eff}} =  \mathcal{O} ( 10 M_p)$ is no longer problematic. Some of these models rely on a large amount of tuning or on the existence of extra dimensions, as 4D dual theories suffer from the same problems as the vanilla model.

In~\cite{DandVrock} we recognised that pGB inflation does not have to have an axion as the inflaton. There are other models which generate a natural inflaton potential, protected from radiative corrections by the same mechanism. In particular, we focussed on compact group structures and showed that one can find models that fit the CMB constraints for a sub-Planckian symmetry breaking scale $f$\footnote{One can also consider non-compact groups such as space time symmetries. In~\cite{Csaki:2014bua} the authors consider broken conformal symmetry and showed that a dilaton inflaton can generate inflation with strictly sub-Planckian scales.}. 

For example, if the pGB field is coupled to external gauge bosons and fermions, a Coleman-Weinberg potential is generated for the inflaton. We demonstrated the general mechanism and gave a specific successful example inspired by the minimal Composite Higgs model MCHM$_5$ ~\cite{Composite-Higgs-orig}.

Here we develop a comprehensive approach to Goldstone Inflaton. In Sec.~\ref{mgsfp}, we give a full analysis of the potentials that can be generated, and motivate that the potential that is uniquely expected to give successful single-field inflation is given by
\begin{equation} \label{Model}
V(\phi) = \Lambda^4 \left( C_\Lambda + \alpha \cos(\phi/f) + \beta \sin^2(\phi/f) \right).
\end{equation}
In Sec.~\ref{consCMB}, we compare its predictions against the CMB data and find that the latter singles out a specific region in the parameter space. We comment on the fine-tuning necessary and show that one obtains a successful model with $f < M_p$ at marginal tuning. 

As the Goldstone inflaton is expected to have non-canonical kinetic terms, we give an analysis of the non-Gaussianity predictions. We show that the current bounds are comfortably evaded. 

In Sec.~\ref{linkUV}, we further explore the region of parameter space that leads to successful inflation. The relations that we find by comparison with the Planck data give information about the form factors that parameterise the UV-theory. 
We comment on the scaling with momentum we expect from theoretical considerations. We finish with an analysis of the UV theory, in which we use QCD-tools to compute the relevant parameters and give a specific example in the approximation of light resonance dominance in Sec.~\ref{lightres}. Finally, in the Appendices we give specific examples of single-field and hybrid inflation coming from Goldstone Inflation.

\section{A successful Coleman-Weinberg potential} 
\label{mgsfp}

Our starting assumption is that the inflaton is a Goldstone boson, coming from the breaking of some global symmetry $G \rightarrow H$. We parameterise the Goldstone bosons using a non-linear sigma model
\begin{equation}
\Sigma(x) = \exp(iT^{\hat a}\phi^{\hat a}(x)/f),
\end{equation}
where $T^{\hat a}$ are the broken $G/H$ generators, $\phi^a(x)$ are the Goldstone fields, and $f$ is the scale of spontaneous symmetry breaking \cite{Callan69}\footnote{Here we assume the CCZW formalism. A different proposal relying on quark seesaw has been made recently (see for instance~\cite{1502.07340} and references therein); however, in this setup the periodicity of the Goldstone field is disguised and therefore we will stick to CCZW.}. Under a $G/H$ symmetry transformation the Goldstone bosons transform via a shift $\phi^a(x) \rightarrow \phi^a + f\alpha^a$, for some transformation parameters $\alpha^a$. This non-linear shift symmetry prevents the Goldstone fields from acquiring a tree-level potential. The inflaton can only get a potential if there are sources of explicit symmetry-breaking that will render $G$ inexact. We list two possibilities:
\begin{enumerate}
\item If the inflaton is a composite object formed of strongly-interacting UV fermions, then explicit fermion mass terms could break the symmetry and give the inflaton a non-zero mass. This would be analogous to the pions of QCD, which acquire a mass from the explicit breaking of chiral symmetry due to the small up and down quark masses.
\item If the inflaton sector has couplings to particles that do not form complete representations of $G$, then loops of these `external' particles will generate a Coleman-Weinberg potential for the inflaton.
\end{enumerate}
Although 1. is an interesting possibility, in this paper we will explore 2., since the Coleman-Weinberg potential can be computed perturbatively (up to coefficients determined by strongly interacting dynamics).

A point worth noting is that, as of yet, we have not fixed the scale at which inflation occurs. The `external' particles relevant to our calculation are those with masses close to, but below the scale of symmetry breaking $\sim f$. If inflation occurs near the TeV scale, we would have to embed the SM gauge group and the heavy quarks into representations of $G$, since these particles would be expected to have the greatest contributions to the inflaton potential (just as in Composite Higgs models). If inflation occurs at the GUT scale $\sim 10^{16}$ GeV, then our lack of knowledge of the high-scale particle spectrum means we can be more open-minded. In the following treatment we leave this question open, considering generic possibilities for the particle content.

That said, we will not consider the contribution from elementary scalars, our prime motivation being the unnaturalness of scalar masses much below the Planck scale. The only scalars appearing in our model will be those coming from the $G/H$ breaking, with masses protected by the non-linear Goldstone symmetry.

We will work through in detail a scenario in which the strong sector has a global $SO(N)$ symmetry which breaks to $SO(N-1)$\footnote{Many of the results of this section generalise straightforwardly to $SU(N) \rightarrow SU(N-1)$.}. This symmetry breaking gives rise to $N-1$ massless Goldstone fields, one linear combination of which will play the role of the inflaton. We will assume that the symmetry-breaking VEV is in the fundamental representation:
\begin{equation}
\Sigma_0 = \langle \Sigma \rangle = \begin{pmatrix} 0 \\ 0 \\ \vdots \\ 1 \end{pmatrix},
\end{equation}
so that
\begin{equation}
\label{sigma}
\Sigma(x) = \exp(iT^{\hat a}\phi^{\hat a}(x)/f)\Sigma_0,
\end{equation}
transforms as a fundamental of $SO(N)$.

\begin{figure}[H]
\begin{tikzpicture}
\draw (0,0) circle [radius=3.5] ;
\draw (0.1,-0.4) circle [radius=2.5] ;
\draw (0.5,-1) circle [radius=1.5] ;
\draw[<-, > = triangle 90] (1.5,0.6) -- (4,1.5) ;
\draw[<-, > = triangle 90] (1, -1.5) -- (5,-0.5) ;
\node at (0,2.5) {$SO(N)$ global symmetry} ;
\node at (0.1,1.2) {$SO(N-1)$ gauged} ;
\node at (0.5,-0.5) {$SO(N-2)$} ;
\node at (0.5,-1) {unbroken} ;
\node at (8,1.5) {$SO(N-1)/SO(N-2)$ massive gauge bosons,} ;
\node at (8.5,1) {couple to the inflaton} ;
\node at (8,-0.5) {$SO(N-2)$ massless gauge bosons,} ;
\node at (8,-1) {do not couple to the inflaton} ;
\end{tikzpicture}
\caption{Subgroups of the global $SO(N)$ symmetry.}
\label{SON}
\end{figure}
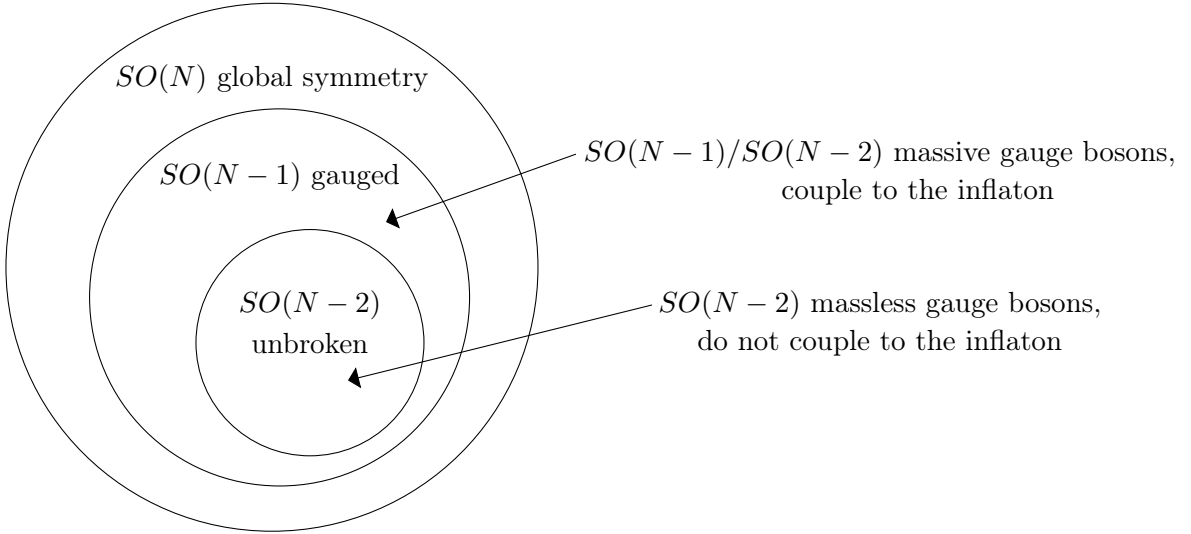

If we take the unbroken symmetry $SO(N-1)$ to be a gauge symmetry, we can gauge away $N-2$ of the Goldstone fields (they give mass to $N-2$ gauge bosons), as we show pictorially in Fig.~\ref{SON}. This will leave us with one physical Goldstone field, which we identify with the inflaton. The same mechanism gives masses to the $W^\pm$ and $Z$ bosons in Composite Higgs models (see for example \cite{Agashe05}, \cite{Gripaios09}). We can gauge a smaller subgroup of $SO(N)$, although this will leave more than one Goldstone degree of freedom. Some possibilities are explored in Appendix \ref{sec:appB}.

We now attempt to write down an effective Lagrangian containing couplings of the Goldstone fields to the $SO(N-1)$ gauge bosons. A useful trick is to take the whole $SO(N)$ global symmetry to be gauged, and only at the end of the calculation setting the unphysical $SO(N)/SO(N-1)$ gauge fields to zero \cite{CHM-reviews}. The most general effective Lagrangian involving couplings between $\Sigma$ and $SO(N)$ gauge bosons, in momentum space and up to quadratic order in the gauge fields, is
\begin{equation}
\label{gauge}
\mathcal L_\mathit{eff} = \frac{1}{2}(P_T)^{\mu\nu}\left[ \Pi_0^A(p^2) \Tr\{A_\mu A_\nu\} + \Pi_1^A(p^2) \Sigma^T A_\mu A_\nu \Sigma \right],
\end{equation}
where $A_\mu = A_\mu^a T^a$ ($a=1,...,N$) are the $SO(N)$ gauge fields, $P_T^{\mu\nu} = \eta^{\mu\nu} - q^\mu q^\nu / q^2$ is the transverse projector, and $\Pi_{0,1}^A(p^2)$ are scale-dependent form factors, parameterising the integrated-out dynamics of the strong sector.

Taking an appropriate choice for the $SO(N)$ generators and expanding out the matrix exponential in \eqref{sigma}, we obtain:
\begin{equation}
\Sigma = \frac{1}{\phi}\begin{pmatrix} \phi^1 \sin(\phi/f)  \\ \vdots \\ \phi^{N-1} \sin(\phi/f) \\ \phi \cos(\phi/f) \end{pmatrix},
\end{equation}
where $\phi = \sqrt{\phi^{\hat a}\phi^{\hat a}}$. With an $SO(N-1)$ gauge transformation we can rotate the $\phi^{\hat a}$ fields along the $\phi^1$ direction, so that
\begin{equation}
\label{rotsigma}
\Sigma = \begin{pmatrix} \sin(\phi/f) \\ \vdots \\ 0 \\ \cos(\phi/f) \end{pmatrix}.
\end{equation}
The remaining $N-2$ degrees of freedom give masses to as many gauge bosons. Expanding out all the terms in \eqref{gauge} and setting the $SO(N)/SO(N-1)$ gauge fields to zero as promised, we obtain:
\begin{equation}
\mathcal L_\mathit{eff} =  \frac{1}{2}(P_T)^{\mu\nu}\left[ \Pi_0^A(p^2) + \frac{1}{2}\Pi_1^A(p^2) \sin^2(\phi/f) \right] A_\mu^{\tilde a} A_\nu^{\tilde a},
\end{equation}
where $A_\mu^{\tilde a}$ are the $SO(N-1)/SO(N-2)$ gauge fields. The remaining (massless) $SO(N-2)$ gauge fields do not couple to the inflaton (See Fig.~\ref{SON}).

Using this Lagrangian we can derive a Coleman-Weinberg potential for the inflaton \cite{Coleman73}:
\begin{equation}
\label{logpotential}
V = \frac{3(N-2)}{2} \int \frac{d^4 p_E}{(2\pi)^4} \log \left[ 1 + \frac{1}{2}\frac{\Pi_1^A}{\Pi_0^A} \sin^2(\phi/f) \right],
\end{equation}
where $p_E^2 = -p^2$ is the Wick-rotated Euclidean momentum. This result can be understood as the sum over the series of diagrams:

\begin{fmffile}{gl1}
\begin{equation}
\label{gl1}
\begin{tikzpicture}[baseline=(current bounding box.center)]
\node{
\fmfframe(1,1)(1,1){
\begin{fmfgraph*}(40,40)
\fmfleft{v1}
\fmfright{v2}
\fmf{dashes}{v1,i1}
\fmf{dashes}{v2,i1}
\fmf{wiggly,right,tension=1}{i1,i1} 
\fmfblob{10}{i1}  
\end{fmfgraph*}
}
};
\end{tikzpicture}
+
\begin{tikzpicture}[baseline=(current bounding box.center)]
\node{
\fmfframe(1,1)(1,1){
\begin{fmfgraph*}(40,40)
\fmfleft{v1,v2}
\fmfright{v3,v4}
\fmfshift{0,20}{v1}
\fmfshift{0,-20}{v2}
\fmfshift{0,20}{v3}
\fmfshift{0,-20}{v4}
\fmf{dashes}{v1,i1}
\fmf{dashes}{v2,i1}
\fmf{dashes}{v3,i2}
\fmf{dashes}{v4,i2}
\fmf{wiggly,left,tension=.5}{i1,i2}
\fmf{wiggly,left,tension=.5}{i2,i1}   
\fmfblob{10}{i1,i2}
\end{fmfgraph*}
}
};
\end{tikzpicture}
+
\begin{tikzpicture}[baseline=(current bounding box.center)]
\node{
\fmfframe(1,1)(1,1){
\begin{fmfgraph*}(40,40)
\fmfsurround{v1,v2,v3,v4,v5,v6}
\fmf{dashes}{v1,i1}
\fmf{dashes}{v2,i1}
\fmf{dashes}{v3,i2}
\fmf{dashes}{v4,i2}
\fmf{dashes}{v5,i3}
\fmf{dashes}{v6,i3}
\fmf{wiggly,tension=.5}{i1,i3}
\fmf{wiggly,tension=.5}{i3,i2}   
\fmf{wiggly,tension=.5}{i2,i1}
\fmfblob{10}{i1,i2,i3}
\end{fmfgraph*}
}
};
\end{tikzpicture}
+ ... ,
\end{equation}
\end{fmffile} 
in which the inflaton field is treated as a constant, classical background. The factor of $3(N-2)$ comes from the 3 degrees of freedom of each of the massive $SO(N-1)/SO(N-2)$ gauge bosons, any of which may propagate around the loop.

As discussed in \cite{CHM-reviews}, $\Pi_1$ can be thought of as an order parameter, which goes to zero in the symmetry-preserving phase at high momenta. Provided the ratio $\Pi_1^A/\Pi_0^A$ decreases fast enough, the integral in \eqref{logpotential} will converge. We can approximate the potential by expanding the logarithm at leading order. This approximation is equivalent to assuming the dominant contribution comes from diagrams with one vertex, and that higher order diagrams are suppressed\footnote{Equivalently
\begin{equation}
\int \frac{d^4 p_E}{(2\pi)^4} \left( \frac{\Pi_1^A}{\Pi_0^A} \right) \gg \int \frac{d^4 p_E}{(2\pi)^4} \frac{1}{2}\left( \frac{\Pi_1^A}{\Pi_0^A} \right)^2 \gg \int \frac{d^4 p_E}{(2\pi)^4} \frac{1}{3}\left( \frac{\Pi_1^A}{\Pi_0^A} \right)^3 \gg ...
\end{equation}
If the form factors behave as described in Section \ref{lightres}, then this is a reasonable approximation.}. This gives
\begin{equation}
V(\phi) = \gamma \sin^2(\phi/f),
\end{equation}
where
\begin{equation}
\gamma = \frac{3(N-2)}{4}\int \frac{d^4 p_E}{(2\pi)^4} \left( \frac{\Pi_1^A}{\Pi_0^A} \right).
\end{equation}

It is worth pointing out that gauge contributions generically lead to a $\sin^2$ type potential at leading order. A $\sin^2$ potential suffers from the same problems as the cosine of Natural Inflation -- it is only flat enough for superplanckian values of $f$.

However, we should also include contributions from external fermions. Just as with the gauge case, the easiest way to write down a general effective Lagrangian is to assume that the fermions are embedded within representations of the full symmetry group $SO(N)$. First we try embedding two Dirac fermions (one left and one right handed) in fundamental $SO(N)$ representations:
\begin{equation}
\Psi_L = \begin{pmatrix} \psi_L \\ \vdots \\ 0 \end{pmatrix}\;,\;\;\Psi_R = \begin{pmatrix} 0 \\ \vdots \\ \psi_R \end{pmatrix}.
\end{equation}
The reader will note that fermions placed anywhere other than the first and $N^\mathit{th}$ entries of these fundamentals will not contribute to the inflaton potential, since they will not couple to the rotated $\Sigma$ \eqref{rotsigma}. We place $\psi_L$ and $\psi_R$ in two separate fundamentals for the sake of generality -- this arrangement will avoid cancellations between terms that would occur if we used the embedding
\begin{equation}
\begin{pmatrix} \psi_L \\ \vdots \\\psi_R \end{pmatrix}.
\end{equation}

The most general $SO(N)$ invariant effective Lagrangian we can write down, up to quadratic order in the fermion fields, is
\begin{equation}
\mathcal L_\mathit{eff} = \sum_{r=L,R} \overline \Psi_r^i \:\slashed p \left[ \Pi_0^r(p)\delta_{ij} + \Pi_1^r(p)\Sigma_i \Sigma_j \right]\Psi_r^j + M(p) \overline \Psi_L^i \Sigma_i \Sigma_j \Psi_R^j + h.c.\;,
\end{equation}
which can be rewritten:
\begin{multline}
\mathcal L_\mathit{eff} = \overline \psi_L \slashed p \left[ \Pi_0^L(p) + \Pi_1^L(p) \sin^2(\phi/f)\right] \psi_L + \overline \psi_R \slashed p \left[ \Pi_0^R(p) + \Pi_1^R(p) \cos^2(\phi/f)\right] \psi_R \\ + M(p)\sin(\phi/f)\cos(\phi/f)\overline\psi_L \psi_R + h.c.
\end{multline}
We can derive the Coleman-Weinberg potential using the formula
\begin{equation}
V = -\frac{1}{2} N_c \int \frac{d^4 p_E}{(2\pi)^4} \Tr \left [ \log \left( \mathcal M \mathcal M^\dagger \right ) \right ],
\end{equation}
which is correct up to terms independent of $\phi$. Here $N_c$ is the number of fermion colours and
\begin{equation}
\mathcal M_{ij} = \frac{\partial^2 \mathcal L}{\partial \psi^i \partial \overline{\psi^j}},
\end{equation}
for all fermions $\psi^i$. We obtain, up to terms independent of $\phi$:
\begin{multline}
\label{potentiallog}
V = -2N_c\int \frac{d^4 p_E}{(2\pi)^4} \log\Bigg[1 + \frac{\Pi_1^L}{\Pi_0^L}\sin^2(\phi/f) + \frac{\Pi_1^R}{\Pi_0^R}\cos^2(\phi/f) + \frac{\Pi_1^L}{\Pi_0^L}\frac{\Pi_1^R}{\Pi_0^R}\sin^2(\phi/f)\cos^2(\phi/f) \\ + \frac{M^2}{p_E^2\Pi_0^L\Pi_0^R}\sin^2(\phi/f)\cos^2(\phi/f) \Bigg].
\end{multline}
The presence of the $\sin^2 \cos^2$ function inside the logarithm is essentially due to the fact that there are loops in which both $\psi_L$ and $\psi_R$ propagate. We have, among other diagrams, the series:
\begin{fmffile}{fl2}
\begin{equation}
\label{fl2}
\begin{tikzpicture}[baseline=(current bounding box.center)]
\node{
\fmfframe(1,1)(1,1){
\begin{fmfgraph*}(40,40)
\fmfleft{v1}
\fmfright{v2}
\fmf{dashes}{v1,i1}
\fmf{dashes}{v2,i2}
\fmf{fermion,left,label=$\psi_L$,tension=0.5}{i1,i2}
\fmf{fermion,left,label=$\psi_R$,tension=0.5}{i2,i1}   
\fmfblob{10}{i1,i2}
\end{fmfgraph*}
}
};
\end{tikzpicture}
+
\begin{tikzpicture}[baseline=(current bounding box.center)]
\node{
\fmfframe(1,1)(1,1){
\begin{fmfgraph*}(40,40)
\fmfsurround{v1,v2,v3,v4}
\fmf{dashes}{v1,i1}
\fmf{dashes}{v2,i2}
\fmf{dashes}{v3,i3}
\fmf{dashes}{v4,i4}
\fmf{fermion,label=$\psi_L$,l.side=left,tension=0.5}{i1,i4}
\fmf{fermion,label=$\psi_R$,l.side=left,tension=0.5}{i4,i3}
\fmf{fermion,label=$\psi_L$,l.side=left,tension=0.5}{i3,i2}   
\fmf{fermion,label=$\psi_R$,l.side=left,tension=0.5}{i2,i1}
\fmfblob{10}{i1,i2,i3,i4}
\end{fmfgraph*}
}
};
\end{tikzpicture}
+ ...
\end{equation}
\end{fmffile}
This series includes diagrams with $2n$ vertices (compare to \eqref{gl1}, which sums over diagrams with $n$ vertices). Thus the resummation leads to a higher order term in the argument of the log. Again we can expand the logarithm at first order to get a potential of the form:
\begin{equation}
V(\phi) = \alpha \sin^2(\phi/f) + \beta \sin^2(\phi/f) \cos^2(\phi/f).
\end{equation}
This potential has a very flat region for $\alpha \simeq \beta$, the flat region being a maximum (minimum) for $\beta>0$ ($\beta<0$). For realistic inflation we require the flat region to be a local maximum, so that the inflaton can roll slowly down the potential. However, since we expect the $\Pi_0$ form factors to be positive (see, for example \cite{Pomarol12}), the expansion of the log gives a \emph{negative} value for $\beta$\footnote{Note that the $(\Pi_1^L\Pi_1^R)/(\Pi_0^L\Pi_0^R)$ term cancels other terms at next order in the expansion, so does not contribute to the potential.}. The gauge contribution -- being of the form $\sin^2(\phi/f)$ -- will not help matters.

Therefore we turn to the next simplest option: embedding the fermions in \emph{spinorial} representations of $SO(N)$. Spinors of $SO(N)$, for odd $N$, have the same number of components as spinors of $SO(N-1)$. The extra gamma matrix $\Gamma^N$ is the chiral matrix, which in the Weyl representation is the only diagonal gamma matrix. Spinors of $SO(N)$ are built from two spinors of $SO(N-2)$ in the same way that Dirac spinors are constructed using two Weyl spinors. We denote these $SO(N-2)$ spinors $\chi_{L,R}$, and embed the fermions as follows:
\begin{equation}
\chi_{L,R} = \begin{pmatrix} \psi_{L,R} \\ 0 \\ \vdots \end{pmatrix},
\end{equation}
and construct the full $SO(N)$ spinors thus:
\begin{equation}
\Psi_L = \begin{pmatrix} \chi_L \\ 0 \end{pmatrix}\;,\;\;\Psi_R = \begin{pmatrix} 0 \\ \chi_R \end{pmatrix}.
\end{equation}
This embedding is chosen so as to ultimately give a coupling between $\psi_L$ and $\psi_R$ -- other embeddings that achieve this will lead to the same eventual result. The $SO(N)$ invariant effective Lagrangian takes the form
\begin{equation}
\label{fermlag}
\mathcal L_\mathit{eff} = \sum_{r=L,R} \overline \Psi_r^i \:\slashed p \left[ \Pi_0^r(p)\delta_{ij} + \Pi_1^r(p)\Gamma^a_{ij} \Sigma^a \right]\Psi_r^j + M(p) \overline \Psi_L^i \Gamma^a_{ij} \Sigma^a \Psi_R^j + h.c.\;,
\end{equation}
where $\Gamma^a$ are the Gamma matrices of $SO(N)$. If we take
\begin{equation}
\Gamma^1 = \begin{pmatrix} 0 & I \\ I & 0 \end{pmatrix}\;,\;\;\Gamma^N = \begin{pmatrix} I & 0 \\ 0 & -I \end{pmatrix}\;
\end{equation}
this can be expanded to give:
\begin{multline}
\label{effL}
\mathcal L_\mathit{eff} = \overline \psi_L \slashed p \left[ \Pi_0^L(p) + \Pi_1^L(p) \cos(\phi/f)\right] \psi_L + \overline \psi_R \slashed p \left[ \Pi_0^R(p) - \Pi_1^R(p) \cos(\phi/f)\right] \psi_R \\ + M(p)\sin(\phi/f)\overline\psi_L \psi_R + h.c.
\end{multline}
Combined with the gauge contribution, this will lead to the potential:
\begin{equation}
\label{potential}
V(\phi) = \alpha \cos(\phi/f) + \beta \sin^2(\phi/f),
\end{equation}
where
\begin{equation}
\alpha = -2N_c \int \frac{d^4 p_E}{(2\pi)^4} \left(\frac{\Pi_1^L}{\Pi_0^L}-\frac{\Pi_1^R}{\Pi_0^R}\right) \;,\;\;\beta = \int \frac{d^4 p_E}{(2\pi)^4} \left( \frac{3(N-2)}{4}\frac{\Pi_1^A}{\Pi_0^A}-2N_c\frac{M^2}{p_E^2\Pi_0^L\Pi_0^R} \right).
\end{equation}
This potential has a flat maximum for $\alpha \simeq 2\beta$, $\beta > 0$. The gauge contribution can now give us a positive value for $\beta$. Thus, for a region of parameter space, this is a viable inflationary potential.

Including more fermions in our model will lead to a wider class of diagrams contributing to the Coleman-Weinberg potential. If we expand consistently to first order in $\Pi_1/\Pi_0$ and $(M/\Pi_0)^2$ however, the only terms that appear at leading order will be those coming from diagrams in which only a single fermion, or an alternating pair of fermions, propagates around the loop. Equation \eqref{potential} will therefore be the generic leading order result, although the coefficients will be modified. In particular, $\alpha$ will be given generally by
\begin{equation}
\alpha = -2N_c \int \frac{d^4 p_E}{(2\pi)^4} \left(\sum_i (-1)^{a_i}\frac{\Pi_1^i}{\Pi_0^i}\right),
\end{equation}
where $a_i=0$ if $\psi_i$ is embedded in the upper half of an $SO(N)$ spinor, and $a_i=1$ if $\psi_i$ is embedded in the lower half.

We should also consider whether including NLO terms in the log expansion changes any of the above conclusions. Assuming that the log expansion is valid, we expect the NLO terms to be suppressed. A small $\sin^4(\phi/f)$ or $\cos(\phi/f) \sin^2(\phi/f)$ addition to the potential will only have the effect of changing slightly the conditions on the coefficients. For example, the potential
\begin{equation}
V(\phi) = \alpha \cos(\phi/f) + \beta \sin^2(\phi/f) + \delta \cos(\phi/f) \sin^2(\phi/f),
\end{equation}
has the flatness condition $\alpha = 2(\beta + \delta)$.

To satisfy the phenomenological constraint that the inflaton potential should be zero at its minimum $V(\phi_\mathit{min}) = 0$, we now insert a constant term $C_\Lambda$ by hand:
\begin{equation}\label{model}
V(\phi) = C_\Lambda + \alpha \cos(\phi/f) + \beta \sin^2(\phi/f).
\end{equation}
In this case, $C_\Lambda = \alpha$. As conventional when writing inflaton potentials we may factor out a scale $\Lambda^4$ to obtain \eqref{Model}, with a redefinition of the coefficients $\alpha$ and $\beta$.

The result that fermions in \emph{fundamental} representations cannot induce a satisfactory inflation potential holds generically for any group, for precisely the reasons outlined above. It is for this reason that we did not consider $SU(N)$ symmetries, since the only single-index representations of $SU(N)$ are fundamental (or anti-fundamental) representations. Embedding fermions in spinorial representations will generally lead, at first order, to a potential of the form \eqref{potential}. Since spinorial representations only exist in $SO(N)$, we conclude that an $SO(N)$ symmetry of the strong sector is the simplest and most natural way to generate a realistic inflaton potential.
Isomorphisms such as $SO( 6 ) \simeq SU(4)$ and $SO(4) \simeq SU(2) \times SU(2)$ allows us to extend this result a little further. For example, embedding fermions in a vector of $SO(4)$ should lead to the same result as fermions embedded in a $(2,2)$ of $SU(2) \times SU(2)$.

\section{Constraints from Inflation}
\label{consCMB}

After our discussion of the general structure of the inflaton potential, let us discuss the restrictions coming from inflation. We list some potentials that can give rise to inflation in Table~\ref{table}.

We parameterise the flatness of the potential as usual in the slow roll approximation (SRA). That is, we require
$\epsilon \ll 1$ and $\eta \ll 1$, where $\epsilon$ and $\eta$ are here given by 
\bea \epsilon = \frac{M_p^2}{2} \left( \frac{  V'(\phi)}{V(\phi)}\right)^2  \textrm{  and   } \eta = M_p^2 \frac{V''(\phi)}{V(\phi) } \ .
\eea

To simplify our expressions, in this section we work in units of reduced Planck mass  $ M_p$; that is, we will rescale our parameters $\phi \rightarrow \frac{\phi}{M_p}$   and  $f \rightarrow \frac{f}{M_p}$.

The number of e-foldings in the slow-roll approximation is then given by 
\bea N = \frac{1}{ \sqrt{2}} \int_{\phi_E}^{\phi_I} \frac{1}{\sqrt{\epsilon}} \eea
where $\phi_E$ is fixed as the field value for which either $\epsilon = 1$ or $\eta = 1$, in other words, the field value for which the SRA breaks down. In our models slow roll breaks down by the second condition. Here and in the following we conservatively choose $N = 60$ for our predictions, and this allows us to find the initial condition for $\phi$. 

We compare the predictions of our model and the CMB data for the spectral tilt and the tensor-to-scalar ratio, which can be expressed in the SRA as
\begin{equation} n_s = 1 + 2 \eta - 6 \epsilon \end{equation}
\begin{equation} r = 16 \epsilon \end{equation}
respectively.

A generic potential for a pseudo-Goldstone boson would contain powers of periodic functions, $c_\phi = \cos \phi /f$ and $s_\phi = \sin \phi /f$, which we parametrize as
\bea
V(\phi) = \Lambda^4 \,  ( C_\Lambda + \sum_n \alpha_n c_\phi^n + \beta_n s_\phi^n )
\eea
The derivatives of this potential are again proportional to the same periodic functions. Roughly speaking, the flatness of the potential can be achieved in two ways. One possibility is setting the argument, $\phi/f$, to be very small (modulo $2\pi$) as in the Natural Inflation scenario. As the fluctuations of the inflaton can be large, this condition typically implies $f \gtrsim M_p$, hence spoiling the predictivity of the model.

Another possibility, and that is what we pursue here, is to look for models with $f < M_p$, which in turn implies that two oscillating terms contribute to the flatness of the potential. This may seem like it would introduce fine-tuning in the model, but in the next section we quantify that tuning, finding it is milder than e.g. Supersymmetry with TeV scale superpartners.   

Note that different models are equivalent from a cosmological perspective and can be transformed into each other by a rotation in parameter space. We list these redefinitions of the parameters and the cosmological constant in Table~\ref{table} as well.

\begin{center}\label{table}
  \begin{tabular}{ l | l | l | l }
    \hline
    Model & $|\tilde\beta |= |\beta / \alpha|$ & $\beta / | \beta |$ & $C_\Lambda$ (pheno) \\ \hline \hline
    $V = \Lambda^4 \left(C_\Lambda + \alpha \cos\frac{\phi}{f} + \beta \sin \frac{\phi}{f}\right) $ &  Like vanilla NI: no   & $+/-$ & $C_\Lambda = \sqrt{\alpha^2 + \beta^2} $\\
    &solution for $ f \leq M_p$& \\ \hline
    {$\!\begin{aligned} 
               V &=  \Lambda^4 \left(C_\Lambda + \alpha \cos\frac{\phi}{f} + \beta \sin^2 \frac{\phi}{f}\right)  \\    
                &=  \Lambda^4 \left(\tilde{C}_\Lambda + \alpha \cos\frac{\phi}{f} - \beta  \cos^2 \frac{\phi}{f}\right)  \end{aligned}$}  & $\lesssim 1/2$  & $+$ & {$\!\begin{aligned} 
               C_\Lambda &=\alpha \\ \tilde{C}_\Lambda &=  \alpha + \beta  \end{aligned}$}  \\ \hline
         {$\!\begin{aligned} 
               V &=  \Lambda^4 \left(C_\Lambda + \alpha \sin^2\frac{\phi}{f} + \beta \sin^2 \frac{\phi}{f}  \cos^2 \frac{\phi}{f}\right)  \\    
                &=  \Lambda^4 \left( \bar{C}_\Lambda - \alpha \cos^2\frac{\phi}{f}  + \beta \sin^2 \frac{\phi}{f}  \cos^2 \frac{\phi}{f}\right)  \\           
                &=  \Lambda^4 \left(C_\Lambda + ( \alpha +\beta ) \sin^2\frac{\phi}{f} - \beta \sin^4 \frac{\phi}{f} \right) \\
                 &=  \Lambda^4 \left(\bar{C}_\Lambda + ( \beta - \alpha ) \cos^2\frac{\phi}{f} -  \beta  \cos^4 \frac{\phi}{f}  \right)  \end{aligned}$} & $\lesssim 1/2$ & $+$ & {$\!\begin{aligned} 
               C_\Lambda &=\alpha \\ \bar{C}_\Lambda &= 2 \alpha  \end{aligned}$} \\    \hline
    \hline
    \end{tabular}
\end{center}

\begin{figure}[ht] 
\centering
  \includegraphics[width= 300pt]{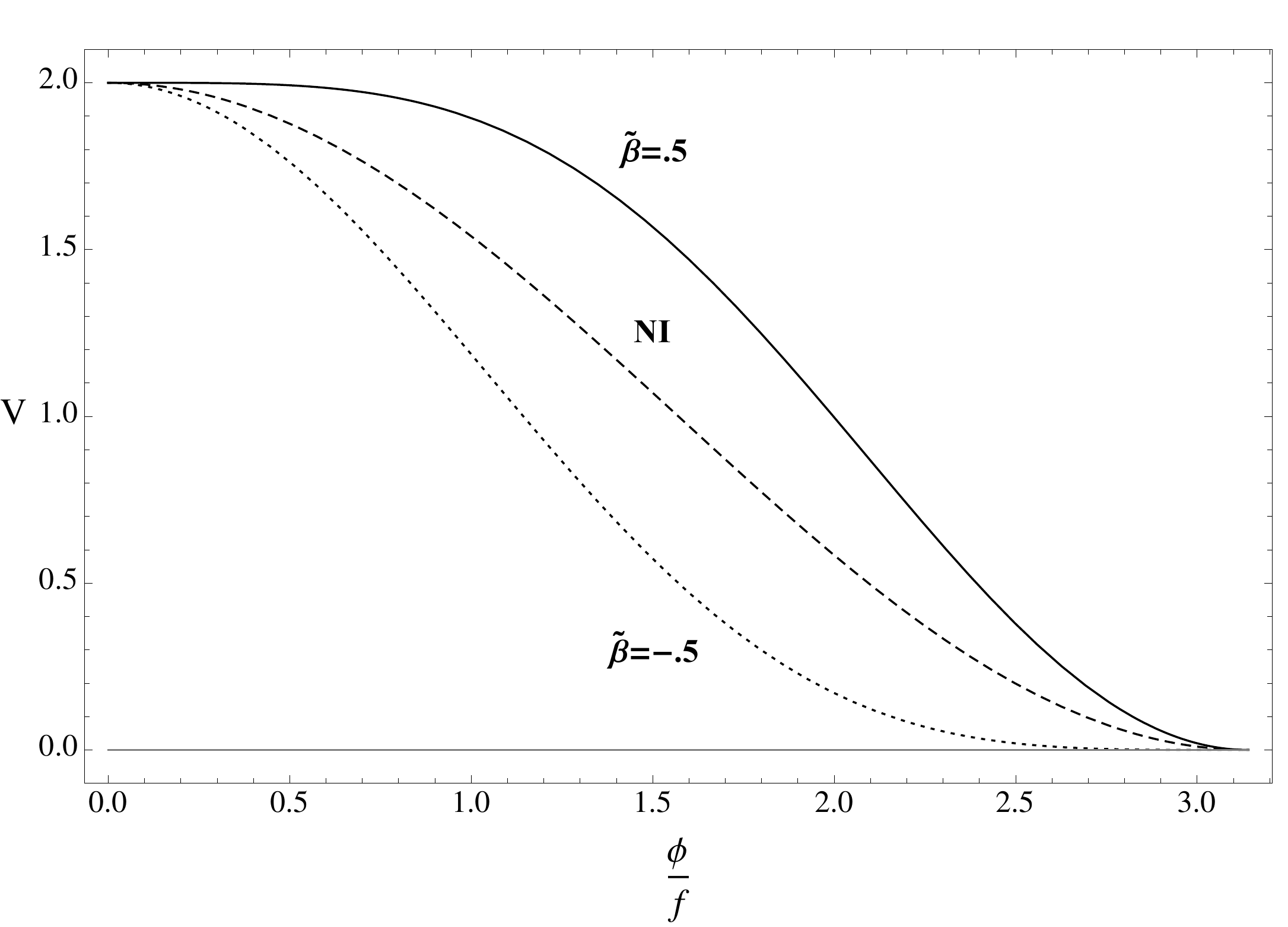}
  \caption{{\it Form of the potential:} Shape of the potential for $\tilde\beta = \pm 1/2$ respectively. Different values will interpolate between these extreme cases. We show the shape of the vanilla NI \eqref{vanilla} for comparison. The height of the potential is normalised to $\Lambda$. }\label{form}
\end{figure}

In the limit that the ratio $\tilde\beta = \beta / \alpha$ is $\pm 1/2$, the potential is exactly flat at the origin and the spectrum is scale-invariant, i.e. $n_s = 1$ as shown in Fig.~\ref{form}.

As the Planck data indicates a small deviation from scale invariance, we expect a small deviation of $\tilde\beta$ with respect to 1/2. We find that the smaller $f$ compared to $M_p$, the closer $\tilde{\beta}$ must be to the values in the table. The deviation $\delta \tilde\beta = 1/2 - \beta $ is then
\begin{equation} 1 \times 10^{-2} \left(\frac{f}{M_p} \right)^2 < \, \delta \tilde\beta \,< 2 \times 10^{-2} \left(\frac{f}{M_p} \right)^2 \end{equation}
for all models in the table, but most importantly the model motivated in the previous section \eqref{model}.
 
\begin{figure}[ht] 
\centering
  \includegraphics[width= 400pt]{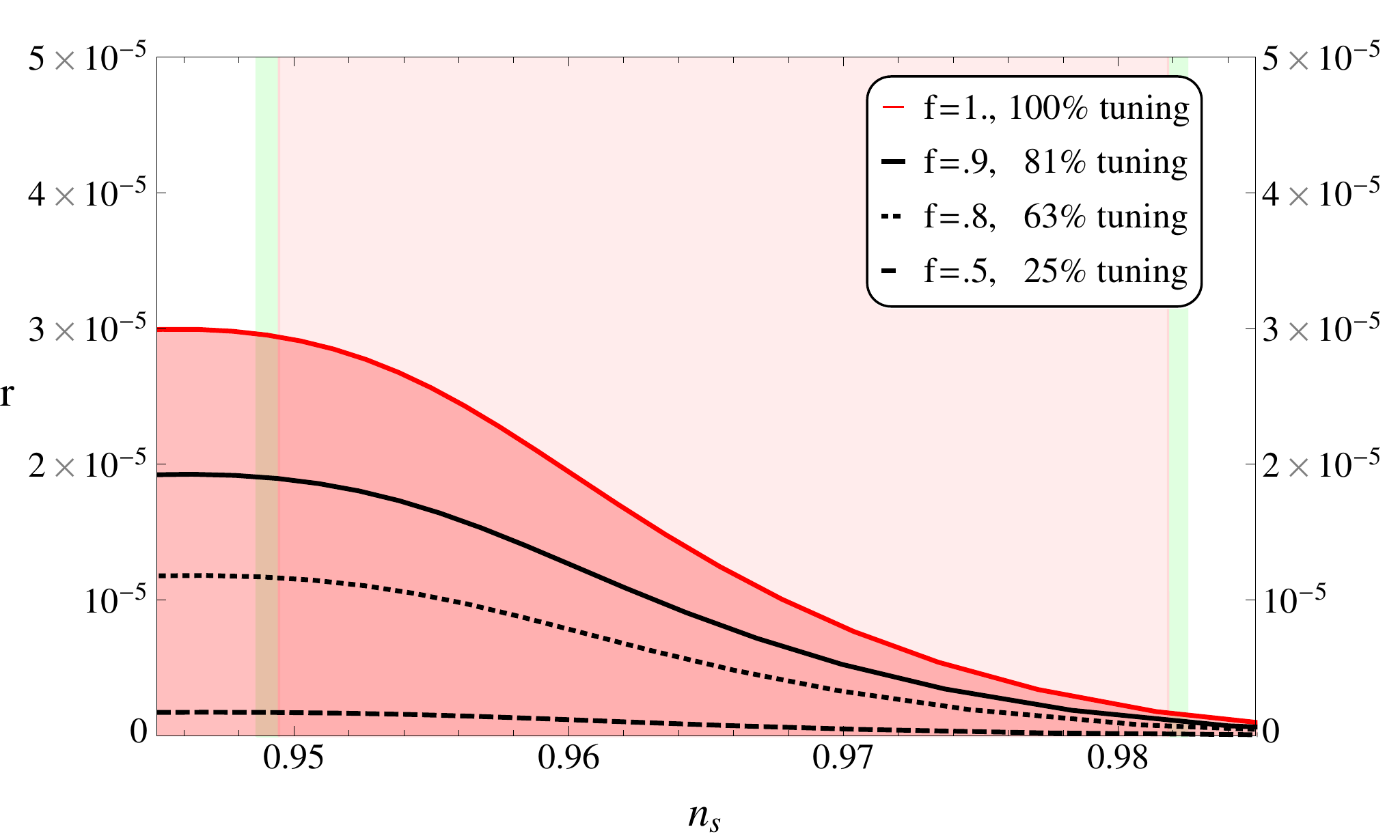}
  \caption{{\it Model predictions:} Parameters $n_s$ and $r$ plotted against the Planck 2015 data~\cite{Planck2015} for the model \eqref{model} for $f = M_p$ (red upper bound). For lower values $f < M_p$, $r \rightarrow 0$ (shaded region). }\label{Nsrplot}
\end{figure}

This is the range of $\tilde\beta$ for which the model is compatible with the Planck data, as we plot in Fig.~\ref{Nsrplot}. for the well motivated example  $V = \Lambda^4 \left(C_\Lambda + \alpha \cos\phi/f+ \beta \sin^2 \phi/f\right) $. Our models predict negligible tensors, so the measurement of $r$ imposes no constraint on $\tilde\beta$. In fact, the tensor to scalar ratio will scale as 
\bea r \propto \left( \frac{f}{M_p} \right)^4 \eea
such that the lower the symmetry breaking scale, the smaller the predicted tensor modes are.

The scale of inflation can be found from the amplitude of the scalar power spectrum, as measured by Planck~\cite{Planck2015}, 
\bea A_s = \frac{\Lambda^4 }{24 \pi^2 M_p^4 \epsilon } = \frac{e^{3.089}}{10^{10}}\eea
where $\epsilon = r /16$ is the first slow roll parameter. For our case this implies
\bea \Lambda_{inf} \approx 10^{15} \left(\frac{f}{M_p} \right) \text{GeV}.\eea 
It is seen that $\Lambda_{inf}$ is expected to be of order of the GUT scale, but can be lower if we allow for tuning. The symmetry breaking should occur before the onset of inflation, and therefore the scale $f$ is expected to lie in the interval $\Lambda_{inf} < f < M_p$. Indeed, from the above relation, it is seen that $\Lambda_{inf} \approx 10^{-3} f$. Lowering the scale $f$ as a result of more tuning thus directly results in lowering the scale inflation; for example, the model predicts $f \approx M_{GUT}\rightarrow \Lambda_{inf} \approx 10^{12} \,\text{GeV}$ for $\delta \tilde\beta \approx 10^{-6}$. 
We will quantify the tuning in the model more precisely in the next section. 

\subsection{Fine-tuning} 

One may note that the specific relationship between $\alpha$ and $\beta$ in the model described above requires one to fine-tune it. Here we quantify the amount of fine-tuning that one will typically expect. 

Defining tuning as is customary in Particle Physics~\cite{finetuning}, we have 
\begin{equation}\label{Delta} \Delta = \left|  \frac{\partial\, \log n_s}{\partial \log \tilde{\beta}} \right| = \left| \frac{\tilde{\beta}}{n_s} \frac{\partial\, n_s}{\partial \tilde{\beta}} \right| \approx \left[ 1.02 - 1.05 \right] \left( \frac{f}{M_p}\right)^{-2} \end{equation}

This relation is not unexpected because for large $f > M_p$ the potential will very flat over a large field range $\Delta\phi$, and this flatness is not sensitive to the specific value of $\tilde\beta$. 
For $f < M_p$ one needs a (partial) cancelation in $\alpha$ and $\beta$, at the cost of fine-tuning.

\begin{figure}[H] 
\centering
  \includegraphics[width= 300pt]{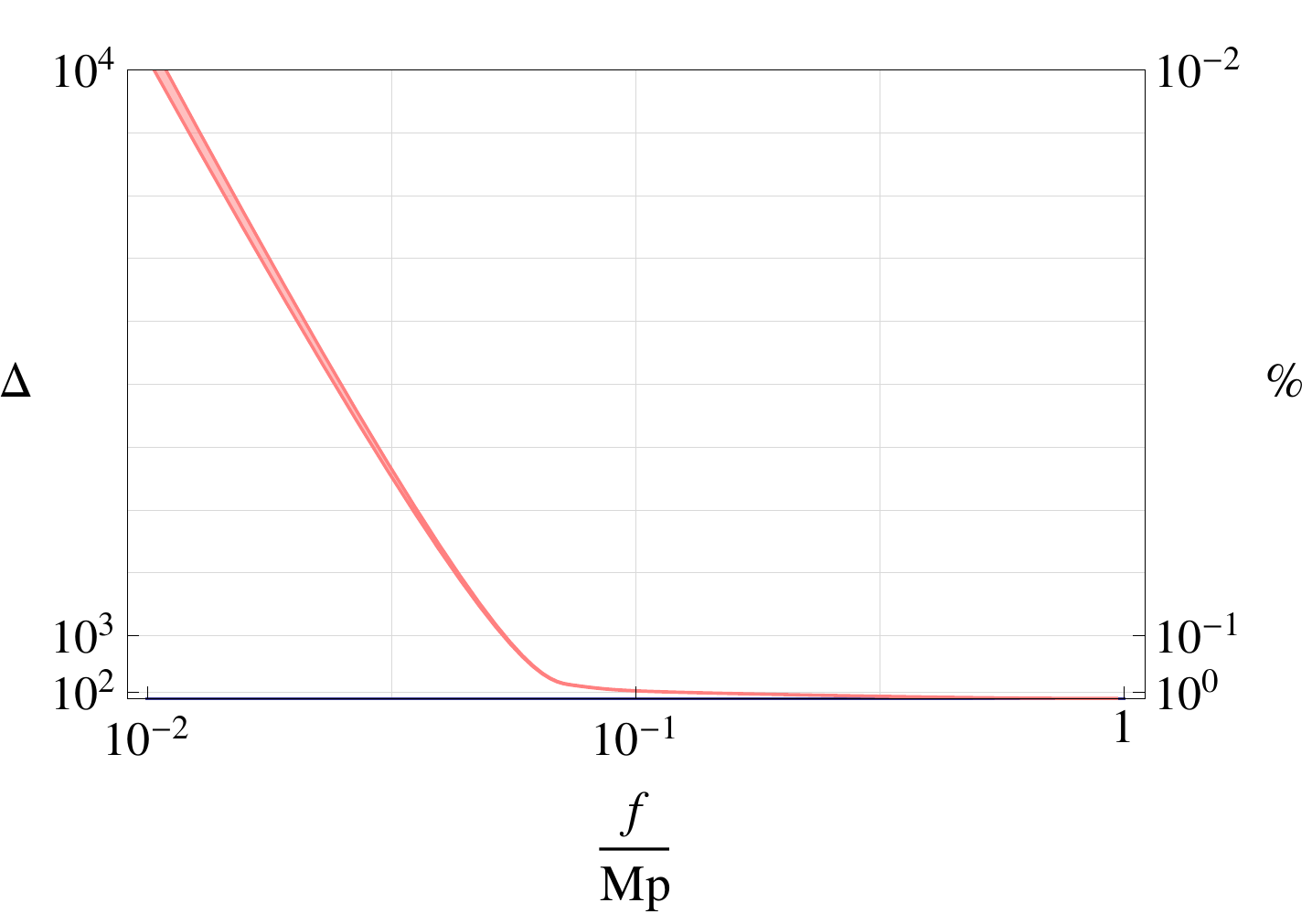}
  \caption{{\it Tuning:} The parameter $\Delta$ as defined above for $V = \Lambda^4 \left(C_\Lambda + \alpha \cos\phi/f+ \beta \sin^2 \phi/f\right) $. Outside of the pink zone the spectral index $n_s$ predicted by the model is incompatible with the Planck data ($n_s < .948$ above the region, $n_s > .982$  below).  }\label{tuning}
\end{figure}

Then we can define the percentage of tuning as 
$$ \text{Percentage tuning} = \frac{100}{\Delta} \% \approx  95 \, \left( \frac{f}{M_p}\right)^{2} \%$$
It is seen in particular that if we take the upper bound $f = M_p$ seriously, the minimal tuning is at $95 \%$. In Fig.~\ref{tuning} we plot the tuning $\Delta$ as defined in \eqref{Delta} for the model at hand, \eqref{Model}. It is seen that for $ M_p / 10\lesssim f < M_p$ one expects no tuning below the percent level. One should note that $ f < 10^{-2} M_p \approx M_{GUT}$ is not expected, as the symmetry breaking pattern should occur before the onset of inflation. 

One can compare this amount of tuning with the one required to avoid the de-stabilization of the electroweak scale in Supersymmetry. For example, stops at 1 TeV require a much worse fine-tuning, at the level of 1\%~\cite{nsusy}.

It is also noteworthy that the tuning necessary in the other models in Table~\ref{table} will be very similar to the tuning in $V = \Lambda^4 \left(C_\Lambda + \alpha \cos\phi/f+ \beta \sin^2 \phi/f\right) $.

\subsection{Non-Gaussianity and its relation to Goldstone scattering}

Even before switching on the Coleman Weinberg potential, Goldstone bosons interact with themselves through higher-order derivative terms. Indeed, consistent with the shift symmetry, one can write terms containing a number of derivatives of the field,
\bea
{\cal L} = \sum_n \frac{c_n}{f^{2 n-4}} X^n \textrm{, with } X = \frac{1}{2} \partial_\mu \Sigma \, \partial^\mu \Sigma^\dagger \label{LX}
\eea

The first order term ($n=1$) is the usual kinetic term, whereas any other term ($n\geqslant 2$) would involve interactions of $2 n$ pions. This expansion is called in the context of Chiral Perturbation Theory~\cite{ChiPT} as order ${\cal O}(p^n)$ in reference to the number of derivatives involved. Goldstone self-interactions appear at order ${\cal O}(p^4)$.

Alongside the Coleman-Weinberg potential we derived in the previous section, the derivative self-interactions are relevant for inflation as well, as a nontrivial speed of sound arises from a non-canonical kinetic term. Specifically, the sound speed is a parameterisation of the difference of the coefficients of the spatial and temporal propagation terms for the Goldstone bosons $\phi$:
\bea \mathcal{L} \ni (\partial_t \phi)^2 - c_s^2 (\partial_i \phi)^2 \eea
This difference arises from higher dimensional kinetic terms $X^n$ and the fact that inflation breaks Lorentz invariance. This can of course already be seen from the metric,
\bea ds^2 = (dt)^2 - a(t)^2 (dx_i)^2 \,\,\,\,\,\,\,\, \rightarrow  \,\,\,\,\,\,\,\, g_{00} \neq g_{ii} \eea

The speed of sound is then given by
\bea c_s^2 = \left(1 + 2 \frac{X \mathcal{L}_{XX}}{\mathcal{L}_X} \right)^{-1} \eea
where $\mathcal{L}_X$ and $\mathcal{L}_{XX}$ denote the first and the second derivative of the Lagrangian with respect to $X$ respectively, and where $c_s$ is expressed in units of the speed of light. It is immediately seen that models with a canonical kinetic term predict $c_s = 1$. 
The background equations of motion can be used to relate coefficients to the Hubble expansion parameter,
\bea\label{Hdot} X \mathcal{L}_X =\dot{H} M_p^2 \approx  c_1 f^4  \eea
To second order, the kinetic term will have the form\footnote{Here we use the expansions $ \mathcal{L} \in (X \mathcal{L}_X  + 2X^2 \mathcal{L}_{XX}) ( \partial_t \phi )^2 /f^4$ and $c_s - 1 \approx \frac{X \mathcal{L}_{XX}}{\mathcal{L}_X}$}
\bea \mathcal{L}_2 = \frac{ M_p^2 \dot{H} + M_p^2 \dot{H} (c_s - 1)}{f^4} (\partial_t \phi)^2 = \frac{M_p^2 \dot{H} c_s}{f^4}  (\partial_t \phi)^2   \eea
Canonically normalising the kinetic term thus implies,
\bea\label{fnorm} f^4 = 2 \dot{H} M_p^2c_s  \eea

\begin{figure}[t] 
\centering
   \includegraphics[width= 300pt]{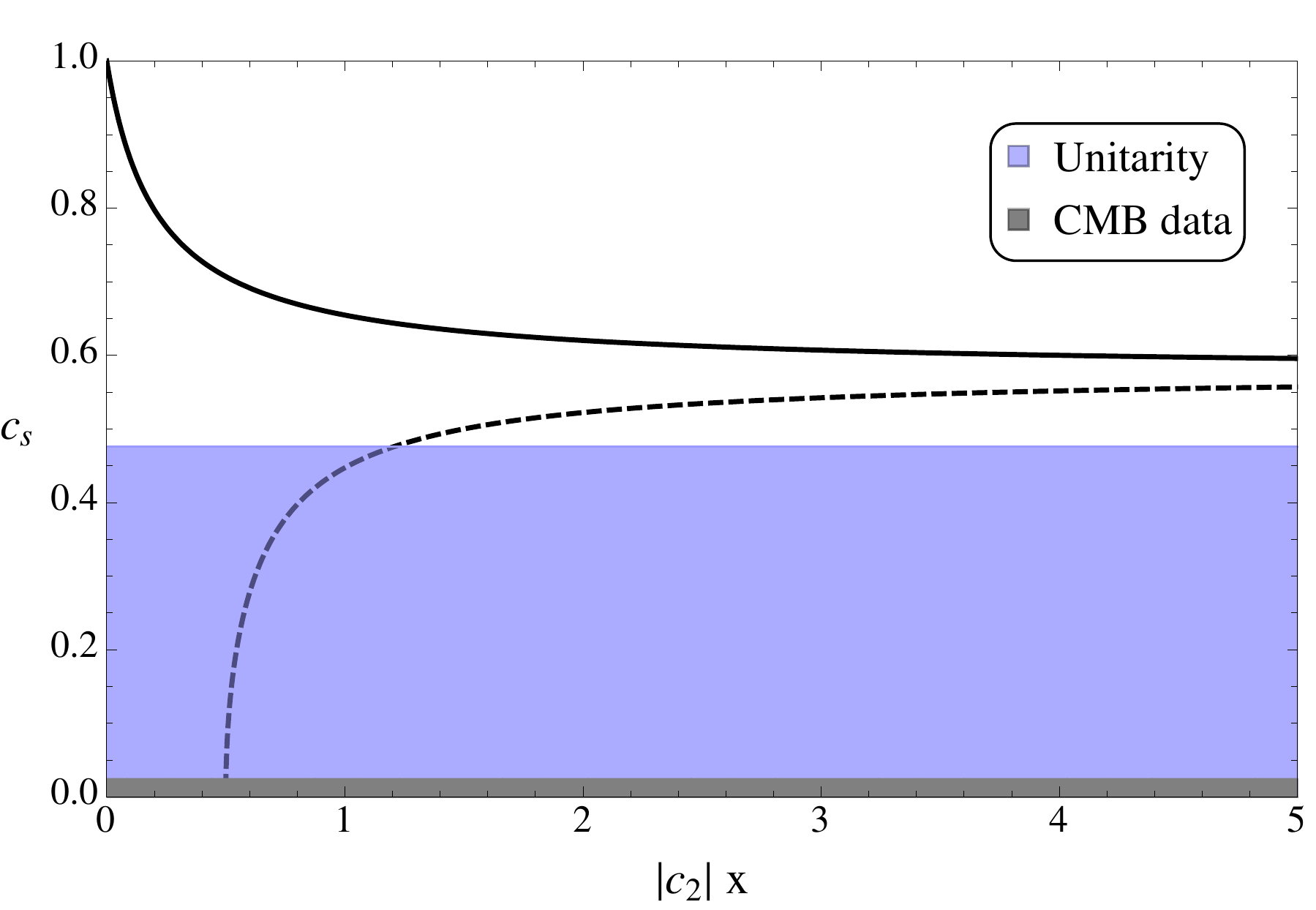}
  \caption{{\it Sound speed:} Predictions for $c_2 \, x$. In dark grey the Planck bound; the shaded region indicates the perturbativity bound. The continuous lines are the predictions for $c_2 >0$, which is relevant for our model as discussed in the text. We also indicate the hypothetical situation $c_2 < 0 $ with dashed curves. It is seen that the prediction approaches the asymptote $c_s = 1 / \sqrt{3}$ for large $c_2$, as expected from \eqref{soundspeedmodel}. }\label{speedofsound}
\end{figure}

These higher order derivatives are also constrained by arguments of unitarity, analyticity and crossing symmetry of Goldstone scattering amplitudes such as shown in Fig.~\ref{speedofsound}, 
\bea
\phi (p_1) \, \phi (p_2) \to \phi (p_3) \, \phi (p_4) \ .
\eea
This scattering amplitude must be a function of the Mandelstam parameters $s$, $t$ and $u$, e.g. $s=(p_1+p_2)^2=(p_3+p_4)^2$.\footnote{In this simplified analysis we have neglected the curvature of space-time. Various issues related to the curvature, and the relevant assumptions one should make to obtain the positivity constraint were discussed in~\cite{curved1} and~\cite{curved2}.}  This amplitude $A(s,t,u)$ must be analytical in the complex $s$ plane, except for branch cuts (due to unitarity) and isolated points (due to the possible exchange of a resonance)~\cite{Manohar}. Unitarity then implies the existence of a branch at some position $s \geqslant s_0$. Similarly, other branch crossings can be obtained by using crossing symmetry. Using these arguments, one can show that the amplitude would be non-analytical for $s> 4 m_{\phi}^2$, where $m_\phi$ is the mass of the pseudo-Goldstone. Moreover, analiticity restricts the dependence of the amplitude on $s$, namely
\bea
\frac{d^2}{d s^2} A(s, t, u) \geqslant 0
\eea
where $s$, $t$ and $u$ are restricted to the physical region, e.g. $s \leqslant 4 m_\phi^2$. 
This translates into bounds for the coefficients of the Lagrangian in \eqref{LX}. At leading order in the Goldstone interactions,
\bea
\mathcal{L}^{(p^4)} =c_2 f^{-4} (\partial_\mu \phi^\dagger \partial^\mu \phi)^2
\eea
the aforementioned conditions lead to a bound for $c_2$. In particular, $c_2$ must be positive and larger than some function of the Goldstone mass.~\footnote{Note that $c_2$ in the case of two- and three-flavour QCD have been computed,  assuming that its dominant contribution comes from vector meson exchange~\cite{GEckerResonances} or with the inclusion of scalar and pseudo-scalar resonances~\cite{1109.3513v2}}

The positivity of $c_2$ constrains possible deviations from the speed of sound in the model with Goldstone inflatons. Indeed, 
\bea \label{soundspeedmodel}
 c_s= \left(1 +2 \frac{2 c_2 X }{f^2+2 c_2 X } \right)^{-1/2} =  \left(1 +2 \frac{2\, c_2 x }{1+2\, c_2  x } \right)^{-1/2}
\eea
Where we have defined the dimensionless parameter $ x = X / f^2 $. As $X \sim p^2$, we expect the effective theory to be valid up to 
\bea\label{alphabound} X \leq (4 \pi f )^2  \,\,\,\,\text{or} \,\,\,\, x \leq (4 \pi)^2
\eea

The current bound by Planck is $c_s > .024$~\cite{Planck2015}. In Fig.~\ref{speedofsound} one can see how for positive $c_2$ the speed of sound is in agreement with Planck for any value of $c_2 \,x$. 

As mentioned above, the sound speed is also constrained by arguments of (perturbative) unitarity. The scale at which violation of perturbative unitarity occurs was computed by Ref.~\cite{1102.5343} (and corrected in~\cite{1407.2621}) from imposing partial wave unitarity in the quartic interaction, and reads, 
\begin{equation} \Lambda_u^4 = \frac{24 \pi}{5} \left( \frac{2 M_p^2 |\dot{H}| c_s^5}{1 - c_s^2} \right) \end{equation}
We are in particular concerned with how $\Lambda_u$ relates to the symmetry breaking scale $f$. If $\Lambda_u <  f$, the action needs a completion below the symmetry breaking scale, possibly in terms of strongly coupled dynamics or new low-energy physics. The effective theory is therefore no longer a good description. One may thus consider a critical sound speed $(c_s)_*$, defined by~\cite{1407.2621} 
\begin{equation} \Lambda_u^4 = \frac{24 \pi}{5} \left( \frac{2 M_p^2 |\dot{H}| (c_s)_*^5}{1 - (c_s)_*^2} \right)= f^4 \end{equation} 
For $c_s > (c_s)_*$ our model predicts $\Lambda_u >  f$. 
Canonically normalising using \eqref{fnorm}, we have
\bea\label{unitcs}  \frac{24 \pi}{5} \left( \frac{  (c_s)_*^4}{1 - (c_s)_*^2} \right)  > 1 \eea
This theoretical lower bound is also shown in Fig.~\ref{speedofsound} for different values of $x$ (subject to \eqref{alphabound}).
One can see how, once axiomatic conditions from Goldstone scattering are imposed, the  inflaton evades both bounds. 

The speed of sound is related to non-Gaussianity by 
\bea f_{NL}^{eq} \sim \frac{1}{c_s^2} \eea
One does not expect significant contributions to non-Gaussianity from the non-derivative terms in the potential, as they will be slow-roll suppressed. 

It is worth noting that a deviation from one in the speed of sound will modify the tensor to scalar ratio 
\bea r = 16 \epsilon c_s \eea
The predictions for $r$ will in this case be lowered, but as the Planck bound is consistent with $r=0$, this is only to the merit of models with a pGB inflaton.

\section{Link to UV models} 
\label{linkUV}

We saw above that the model \eqref{Model} gives inflation compatible with the CMB data for particular relations between the coefficients. Here we discuss what these relations indicate for the UV theory. 

Firstly, we noticed that to have the right shape of the potential, we should require $\beta$ to be positive, that is

\begin{equation}\label{beta>zero} \beta =  \int \frac{d^4 p_E}{(2\pi  \Lambda)^4} \left( \frac{3 (N-2)}{4}\frac{\Pi_1^A}{\Pi_0^A}-2N_c\frac{M^2}{p_E^2\Pi_0^L\Pi_0^R} \right) > 0\end{equation}

Then we saw in Table~{table} that the requirement of a sufficiently flat potential gives the condition $\alpha \approx 2 \beta$, which will give a relation between the form factors of the form

\begin{align} \notag
\alpha &= -2N_c \int \frac{d^4 p_E}{(2\pi \Lambda)^4} \left(\frac{\Pi_1^L}{\Pi_0^L}-\frac{\Pi_1^R}{\Pi_0^R}\right) \\
&= 2   \int \frac{d^4 p_E}{(2\pi \Lambda)^4} \left( \frac{3(N-2)}{4}\frac{\Pi_1^A}{\Pi_0^A}-2N_c\frac{M^2}{p_E^2\Pi_0^L\Pi_0^R} \right)= 2 \beta 
\end{align}

Lastly we have that the phenomenological condition $V(\phi_{min}) = 0$ gives a preferred value of the constant $C_\Lambda$ in terms of the model parameters. In explicit models this will give a condition of the form\footnote{Note that in some models $C_\Lambda$ will be related to $\alpha \pm \beta \approx 3/2\, \alpha$, as is seen in Table~\ref{table}.}

\begin{align} \label{CC}
\alpha = -2 N_c \int \frac{d^4 p_E}{(2 \pi  \Lambda)^4} \left[ \frac{\Pi_1^L}{\Pi_0^L} - \frac{\Pi_1^R}{\Pi_0^R} \right] = C_\Lambda
\end{align}
where $C_\Lambda$ is a cosmological constant during inflation. 

To obtain explicit expressions for the form factors $\Pi_X$ one would need a UV-complete theory. However, using the relations above we can make some general remarks about their large momentum behaviour. 
First, we can use an operator product expansion to find the scaling of $\Pi_1$. This implies that $\Pi_1$ scales as $ \langle \mathcal{O} \rangle / p^{d-2}$, where $\mathcal{O}$ is the lowest operator responsible for the breaking $G \rightarrow H$, with mass dimension $d$. In our case, we expect $\mathcal{O}$ to be a fermion condensate with $d = 6 $. 
Secondly we can require finiteness of the fermion Lagrangian \eqref{fermlag}. 
The scaling of the other form factors can be found by consideration of the kinetic terms in the high momentum limit. We will discuss this in the next section. 
We summarise our conclusions in Table~\ref{table2}. 

\begin{center}\label{table2}
  \begin{tabular}{ l | l | l }
    \hline
    Form factor & Large momentum behaviour & Argument \\ \hline \hline
    $\Pi_1$ & $\sim \langle \mathcal{O} \rangle / p^{d-2}= 1/p^2$ & OPE coupling \\ 
    \hline
    $\Pi_0$ & $\sim p^2  $  & Recovering the bosonic Lagrangian \\ 
    \hline \hline
    $\Pi_1^r$ & $\sim1/p^6$ & OPE coupling \\ 
    \hline
    $\Pi_0^r$ & $\sim p^0 $ & Recovering the fermion Lagrangian  \\ 
    \hline
    $M^r$ & $\sim1/p^2$ & OPE coupling \\ 
    \hline
  \end{tabular}
\end{center}

In the next section we will assume a light resonance connection to derive more specific conclusions in this approximation.

\section{Light resonance connection}
\label{lightres}
In this section we attempt to derive some of the properties of the UV theory, assuming that the integrated-out dynamics is dominated by the lightest resonances of the strong sector.

To simplify what follows, we note that the form factor $M$ in equation \eqref{effL} is `naturally' small in the 't Hooft sense \cite{tHooft80}. This is because in the limit $M \rightarrow 0$ we have an enhanced $U(1)_L \times U(1)_R$ global symmetry under which $\psi_L$ and $\psi_R$ transform with independent phase-rotations. Therefore in the following we will assume that the dominant contributions to $\alpha$ and $\beta$ come from the $\Pi_{0,1}^i$ form factors. Note that this observation makes it very plausible that condition \eqref{beta>zero} is satisfied.

Note that to ensure a convergent behaviour of the form factors at high scales $Q^2$, one would have to introduce an amount of resonances to saturate the Weinberg sum rules. The minimum number of resonances depends on the behaviour of the form factor with $Q$. This behaviour is described in the previous section. Tthe convergence of these form factors in the large-$Q$ regime is not a necessary condition for a generic model of strong interactions, but rather helps on describing the interpolation between the low-energy regime of the theory with an asymptotically free UV theory (provided this is the case). 

Irrespective of these issues of interpolation with the UV behaviour, one can consider a spectral decomposition of the form factor. A common description, valid for a $SU(N)$ gauge sector in the large-$N$ limit is form factors as infinite sums over narrow resonances of the strong dynamics \cite{tHooft74, Witten79}. In the following, we assume that the $\Pi_1^i$ form factors can be well approximated by considering only the contribution from the lightest of these resonances.

We expect that $\Pi_1^i$ has a pole at the mass of the lightest resonance $m_i^2$, and that the residue of this pole is equal to the square of the amplitude to create the resonance from the vacuum. This amplitude, $f_i$, is equivalent to the decay constant of the resonance. This leads us to the following approximation for the fermionic $\Pi_1^i$:
\begin{equation}
\label{Pi1i}
\Pi_1^i(p^2) = \frac{f_i^2}{p^2+m_i^2}.
\end{equation}
In the gauge case, this expression is modified to
\begin{equation}
\label{Pi1A}
\frac{1}{p^2}\Pi_1^A(p^2) = \frac{f^2}{p^2} + \frac{f_A^2}{p^2+m_A^2},
\end{equation}
which now has a pole at $p^2=0$, since the broken $SO(N)/SO(N-1)$ currents can excite the Goldstones from the vacuum \cite{CHM-reviews}.

We approximate the $\Pi_0$ form factors with their tree level values. By inspecting \eqref{gauge} and \eqref{fermlag}, we see that to recover the tree level fermion and gauge Lagrangians we must have $\Pi_0 = 1$ in the fermionic case, and $\Pi_0=p^2/g^2$ in the gauge case, where $g$ is the gauge coupling.

Let us study the minimal model we can construct that leads to successful inflation. We will only need one external fermion -- in this case we take the $\psi_R$ of Sec.~\ref{mgsfp}. Then $\alpha$ and $\beta$ will be given by
\begin{equation}
\label{alphabeta}
\alpha = 2N_c \int \frac{d^4 p}{(2\pi \Lambda)^4} \left( \frac{\Pi_1^R}{\Pi_0^R} \right)\;,\;\;\beta = \frac{3(N-2)}{4} \int \frac{d^4 p}{(2\pi \Lambda)^4} \left( \frac{\Pi_1^A}{\Pi_0^A} \right).
\end{equation}
Now we assume that $\Pi_1^R$ and $\Pi_1^A$ are given respectively by \eqref{Pi1i} and \eqref{Pi1A}. With a single resonance, we cannot guarantee convergence of the integrals in \eqref{alphabeta} -- generally this can be done by introducing more resonances and demanding that the form factors satisfy Weinberg sum rules \cite{Weinberg67, Knecht97}. However we can argue that, since our effective theory is only expected to be valid up to a scale $\Lambda_{UV} = 4\pi f$, we should cut off the momentum integrals at $p^2 = \Lambda_{UV}^2$.

Putting all this together, we find:
\begin{equation}
\alpha = \frac{a}{8\pi^2 \Lambda^4} \int_0^{\Lambda_{UV}^2} dp^2 \frac{p^2 f_R^2}{p^2+m_R^2} = \frac{a f_R^2}{8\pi^2 \Lambda^4} \left[ \Lambda_{UV}^2 - m_R^2 \log\left( \frac{m_R^2+\Lambda_{UV}^2}{m_R^2} \right)\right],
\end{equation}
where $a = 2N_c$, and
\begin{equation}
\beta = \frac{b g^2}{8\pi^2 \Lambda^4} \int_0^{\Lambda_{UV}^2} dp^2 f^2 = \frac{b \: g^2}{8\pi^2}\left[\Lambda_{UV}^2 f^2 + \Lambda_{UV}^2 f_A^2 - f_A^2 m_A^2 \log \left(\frac{m_A^2 + \Lambda_{UV}^2}{m_A^2} \right)\right],
\end{equation}
where $b = 3(N-2)/4$.

The approximate relation $\alpha \simeq 2\beta$ then implies a relationship between the parameters of the UV theory. If we demand that the quadratic cutoff dependence cancels, we obtain the relation
\begin{equation}
\label{f_relation}
a f_R^2 = 2 b g^2 (f^2 + f_A^2),
\end{equation}
and
\begin{equation}
\label{m_relation}
a f_R^2 m_R^2 \log \left( \frac{m_R^2+\Lambda_{UV}^2}{m_R^2} \right) = 2b g^2 f_A^2 m_A^2 \log \left( \frac{m_A^2+\Lambda_{UV}^2}{m_A^2} \right).
\end{equation}
Inserting \eqref{f_relation} into \eqref{m_relation} we obtain
\begin{equation}
\frac{2bg^2 f_A^2}{a f_R^2} = \frac{f_A^2}{f^2+f_A^2} = \frac{m_R^2\log[(m_R^2+\Lambda_{UV}^2)/m_R^2]}{m_A^2\log[(m_A^2+\Lambda_{UV}^2)/m_A^2]},
\end{equation}
which implies that $m_R < m_A$.

If $f_A \gg f$, one finds that $m_R \simeq m_A$, i.e. there would be a degeneracy between fermionic and bosonic resonances. Note that this condition will be satisfied no matter the scale factor between $\alpha$ and $\beta$ is, as long as they are proportional, $\alpha \propto \beta$. This kind of {\it mass-matching} situation~\cite{massmatch} where resonances from different sectors acquire the same mass is reminiscent of what had been found in trying to build successful Technicolor models, namely Cured Higgsless~\cite{cured} and Holographic Technicolor~\cite{HTC} models.  

\section{Discussion and Conclusions}

The framework of slow-roll inflation has been corroborated to a good precision by  the Planck data. This framework, however, suffers from an {\it inflationary hierarchy problem}, namely the strain of providing sufficient inflation while still satisfying the amplitude of the CMB anisotropy measurements. This balancing act requires a specific type of potential, with a width  much larger than its height. 

This tuning is generically unstable unless some symmetry protects the form of the potential. In this paper we explored the idea that this potential could be related to the inflaton as a Goldstone boson, arising from the spontaneous breaking of a global symmetry. 

Another issue for inflationary potentials, including Goldstone Inflation, is that they are only effective descriptions of the inflaton physics. With the inflationary scale relatively close to the scale of Quantum Gravity, one expects higher-dimensional corrections to the inflationary potential. These corrections would de-stabilise the inflationary potential unless the model is small-field~\cite{HDOs}. In other words, as the inflaton field value approaches $M_p$ the Effective Theory approach breaks down.

We found out that in Goldstone Inflation a predictive effective theory is indeed possible, and it leads to specific predictions. For example, in single-field inflation, we computed the most general Coleman-Weinberg inflaton potential and learnt that {\it 1.)} Only the breaking of $SO(N)$ groups provide successful inflation and {\it 2.)} fermionic and bosonic contributions to the potential must be present and {\it 3.)} for fermions in single-index representations, a successful inflaton potential is given uniquely by $V=\Lambda^4(C_\Lambda + \alpha \cos(\phi/f) + \beta \sin^2(\phi/f) )$, with $\alpha \approx 2\beta$. When linking to UV completions of Goldstone Inflation, we have been able to show how relations among the fermionic and bosonic resonances are linked to the flatness of the potential.

As we have developed a specific model for inflation, we were able to address the amount of tuning required to make it work, and found that it is not dramatic. Indeed, we found that the tuning is milder than that found in Supersymmetric models nowadays.

Another advantage of this framework is the ability to examine the higher-order derivative terms in the Goldstone Lagrangian from several different points of view: modifications of the CMB speed of sound, constraints from unitarity and also axiomatic principles from Goldstone scattering.

We have presented results in a rather generic fashion and for single-field inflation, and delegated to the appendices a discussion of a specific model of single-field inflation, and few examples of hybrid inflation which originate from this framework. 

There are other aspects of Goldstone Inflation which deserve further study. For example, in these models, hybrid inflation and reheating are quite predictive as the inflaton and waterfall fields come from the same object and naturally the inflaton can decay to other, lighter pseudo-Goldstones. Moreover, there may be interesting features of the phase transition causing the spontaneous breaking of the global symmetry, which we plan to investigate.

\section*{Acknowledgements}
We would like to thank Juanjo Sanz-Cillero for discussion on the Goldstone boson scattering, and Ewan Tarrant for explaing aspects of reheating. This work  is supported by the Science Technology and Facilities Council (STFC) under grant number ST/L000504/1. 
\appendix

\section{Successful patterns of breaking: an example of single field}
\label{sec:app}

The simplest instance of the general model outlined in Section~\ref{mgsfp} takes the global symmetry of the strong sector to be $SO(3)$, breaking to $SO(2)$.\footnote{This coset was also studied in the context of inflation in~\cite{KinneyMahan}.} This gives rise to two Goldstone bosons, one of which is eaten when we gauge the remaining $SO(2)$ symmetry. We parameterise the Goldstones via:
\begin{equation}
\label{SigmaA}
\Sigma(x) = \exp(iT^{\hat a}\phi^{\hat a}/f)\Sigma_0,
\end{equation}
with $\hat a = 1,2$. We can take the generators of $SO(3)$ to be
\begin{equation}
T^1 = \frac{i}{\sqrt 2} \begin{pmatrix} 0 & 0 & 0 \\ 0 & 0 & -1 \\ 0 & 1 & 0 \end{pmatrix}\;,\;\;T^2 = \frac{i}{\sqrt 2} \begin{pmatrix} 0 & 0 & 1 \\ 0 & 0 & 0 \\ -1 & 0 & 0 \end{pmatrix}\;,\;\;T^3 = \frac{i}{\sqrt 2} \begin{pmatrix} 0 & -1 & 0 \\ 1 & 0 & 0 \\ 0 & 0 & 0 \end{pmatrix}.
\end{equation}
The broken generators satisfy $T^{\hat a}\Sigma_0 \neq 0$. If, following Sec.~\ref{mgsfp}, we take $\Sigma_0$ to be
\begin{equation}
\Sigma_0 = \begin{pmatrix} 0 \\ 0 \\ 1 \end{pmatrix},
\end{equation}
then $T^1$ and $T^2$ are the broken generators. $T^3$ remains unbroken, and will generate the $SO(2)$ gauge symmetry. A suitable gauge transformation then allows us to set $\phi^1=\phi$, $\phi^2=0$, and we can write
\begin{equation}
\Sigma = \begin{pmatrix} \sin(\phi/f) \\ 0 \\ \cos(\phi/f) \end{pmatrix}.
\end{equation}
Following \eqref{gauge} the effective Lagrangian for the $SO(2)$ gauge boson is
\begin{equation}
\mathcal L_\mathit{eff} = \frac{1}{2}(P_T)^{\mu\nu}\left[ \Pi_0^A(p^2) A_\mu^3 A_\nu^3 \Tr\{T^3 T^3\} + \Pi_1^A(p^2) A_\mu^3 A_\nu^3 \Sigma^T T^3 T^3 \Sigma \right],
\end{equation}
\begin{equation}
= \frac{1}{2}(P_T)^{\mu\nu}\left[ \Pi_0^A(p^2) + \frac{1}{2}\Pi_1^A(p^2) \sin^2(\phi/f) \right]A_\mu^3 A_\nu^3.
\end{equation}
This leads to the Coleman-Weinberg potential
\begin{equation}
V = \frac{3}{2} \int \frac{d^4 p}{(2\pi)^4} \log \left[1 + \frac{1}{2}\frac{\Pi_1^A}{\Pi_0^A}\sin^2(\phi/f)\right].
\end{equation}

Now we embed a fermion in an $SO(3)$ spinor:
\begin{equation}
\Psi_L = \begin{pmatrix} \psi_L \\ 0 \end{pmatrix}.
\end{equation}
The gamma matrices of $SO(3)$ can be taken to be the Pauli matrices $\sigma^a$. Thus the most general effective Lagrangian for the fermion is
\begin{equation}
\mathcal L_\mathit{eff} = \overline \Psi_L \slashed p \left[ \Pi_0^L(p) + \Pi_1^L(p) \sigma^a \Sigma^a\right] \Psi_L.
\end{equation}
We find that
\begin{equation}
\sigma^a \Sigma^a = \begin{pmatrix} \cos(\phi/f) & \sin(\phi/f) \\ \sin(\phi/f) & -\cos(\phi/f) \end{pmatrix},
\end{equation}
so
\begin{equation}
\label{fermionlag}
\mathcal L_\mathit{eff} = \overline \Psi_L \slashed p \left[ \Pi_0^L(p) + \Pi_1^L(p) \cos(\phi/f) \right] \Psi_L,
\end{equation}
from which we derive the Coleman-Weinberg potential:
\begin{equation}
V = -2N_c \int \frac{d^4 p}{(2\pi)^4} \log \left[1 + \frac{\Pi_1^L}{\Pi_0^L}\cos(\phi/f) \right].
\end{equation}
Combining both gauge and fermion contributions, and expanding the logs at first order, we obtain
\begin{equation}
V(\phi) = \alpha \cos(\phi/f) + \beta \sin^2(\phi/f),
\end{equation}
where
\begin{equation}
\alpha = -2N_c \int \frac{d^4 p}{(2\pi)^4} \left( \frac {\Pi_1^L}{\Pi_0^L} \right)\;,\;\;\beta = \frac{3}{4}\int \frac{d^4 p}{(2\pi)^4} \left( \frac {\Pi_1^A}{\Pi_0^A} \right).
\end{equation}

\section{Successful patterns of breaking: an example of hybrid inflation}
\label{sec:appB}

We can also construct models in which more than one physical Goldstone degree of freedom is left in the spectrum. This can be done by only gauging a subgroup of the unbroken $SO(N-1)$ symmetry. Let us look briefly at a simple example of such a model, in which we take the global symmetry breaking to be $SO(5) \rightarrow SO(4)$. In such a case we have four Goldstone bosons, and $\Sigma$ is given by
\begin{equation}
\Sigma = \frac{\sin(\phi/f)}{\phi} \begin{pmatrix} \phi^1 \\ \phi^2 \\ \phi^3 \\ \phi^4 \\ \phi \cot(\phi/f)\end{pmatrix},
\end{equation}
where we have $\phi = \sqrt{\phi^{\hat a} \phi^{\hat a}}$, as before.

If we gauge only $SO(2) \in SO(4)$, taking for instance the gauged generator to be
\begin{equation}
T_g^1 = \frac{i}{\sqrt 2} \begin{pmatrix} 0 & 0 & 0 & -1 & 0 \\ 0 & 0 & 0 & 0 & 0 \\ 0 & 0 & 0 & 0 & 0 \\ 1 & 0 & 0 & 0 & 0 \\ 0 & 0 & 0 & 0 & 0 \end{pmatrix},
\end{equation}
then the gauge freedom allows us to set $\phi^4 = 0$.

Following the same steps as before, the effective Lagrangian for the gauge field will be
\begin{equation}
\mathcal L_\mathit{eff} = \frac{1}{2}(P_T)^{\mu\nu}\left[ \Pi_0^A(p^2) + \frac{1}{2}\Pi_1^A(p^2) \left(\frac{\phi^1}{\phi}\right)^2 \sin^2(\phi/f) \right]A_\mu A_\nu.
\end{equation}

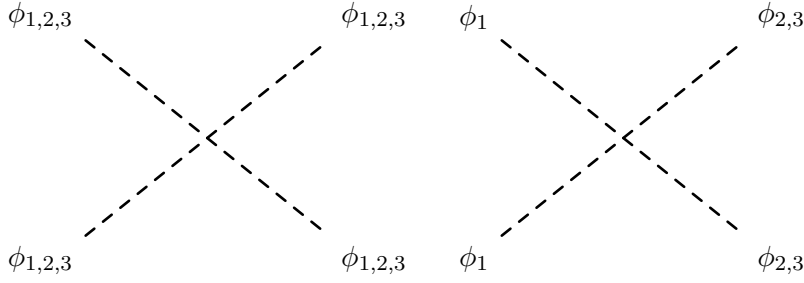
\begin{figure}
\centering
\begin{fmffile}{goldstones}
\begin{tikzpicture}[baseline=(current bounding box.center)]
\node{
\fmfframe(1,1)(1,1){
\begin{fmfgraph*}(40,40)
\fmfleft{v1,v2}
\fmfright{v3,v4}
\fmfshift{0,20}{v1}
\fmfshift{0,-20}{v2}
\fmfshift{0,20}{v3}
\fmfshift{0,-20}{v4}
\fmflabel{$\phi_{1,2,3}$}{v1}
\fmflabel{$\phi_{1,2,3}$}{v2}
\fmflabel{$\phi_{1,2,3}$}{v3}
\fmflabel{$\phi_{1,2,3}$}{v4}
\fmf{dashes}{v1,i1}
\fmf{dashes}{v2,i1}
\fmf{dashes}{i1,v3}
\fmf{dashes}{i1,v4}  
\end{fmfgraph*}
}
};
\end{tikzpicture}
\;\;\;\;\;\;
\begin{tikzpicture}[baseline=(current bounding box.center)]
\node{
\fmfframe(1,1)(1,1){
\begin{fmfgraph*}(40,40)
\fmfleft{v1,v2}
\fmfright{v3,v4}
\fmfshift{0,20}{v1}
\fmfshift{0,-20}{v2}
\fmfshift{0,20}{v3}
\fmfshift{0,-20}{v4}
\fmflabel{$\phi_{1}$}{v1}
\fmflabel{$\phi_{1}$}{v2}
\fmflabel{$\phi_{2,3}$}{v3}
\fmflabel{$\phi_{2,3}$}{v4}
\fmf{dashes}{v1,i1}
\fmf{dashes}{v2,i1}
\fmf{dashes}{i1,v3}
\fmf{dashes}{i1,v4}  
\end{fmfgraph*}
}
};
\end{tikzpicture}
\end{fmffile}
\caption{Goldstone quartic interactions}
\label{quartics}
\end{figure}

If we, as in Appendix~\ref{sec:app}, consider the contribution from a single left-handed fermion, now embedded in an $SO(5)$ spinor like so:
\begin{equation}
\Psi_L = \begin{pmatrix} \psi_L \\ 0 \\ 0 \\ 0 \end{pmatrix},
\end{equation}
then in fact the effective fermion Lagrangian will still be given by \eqref{fermionlag}. Thus the Coleman-Weinberg potential will be given by
\begin{equation}
V(\phi) = \alpha \cos(\phi/f) + \beta \left(\frac{\phi_1}{\phi} \right)^2 \sin^2(\phi/f),
\end{equation}
with $\alpha$ and $\beta$ given by
\begin{equation}
\alpha = -2N_c \int \frac{d^4 p}{(2\pi)^4} \left( \frac {\Pi_1^L}{\Pi_0^L} \right)\;,\;\;\beta = \frac{3}{4}\int \frac{d^4 p}{(2\pi)^4} \left( \frac {\Pi_1^A}{\Pi_0^A} \right).
\end{equation}
If we expand the trigonometric functions for small field excursions, we obtain, up to constant terms:
\begin{multline}
\label{hybridpotential}
V(\phi_1,\phi_2,\phi_3) = \frac{1}{f^2}\left( \beta-\frac{\alpha}{2}\right) \phi_1^2 - \frac{\alpha}{2f^2}\left( \phi_2^2 + \phi_3^2 \right) + \frac{1}{f^4}\left(\frac{\alpha}{24} - \frac{\beta}{3} \right) \phi_1^4 \\ + \frac{\alpha}{24f^4}\left( \phi_2^4 + \phi_3^4 \right) + \frac{1}{f^4}\left( \frac{\alpha}{12} - \frac{\beta}{3} \right) \left( \phi_1^2\phi_2^2 + \phi_1^2\phi_3^2 \right) + \frac{\alpha}{12f^4}\phi_2^2\phi_3^2 + \mathcal O\left(\frac{\phi^6}{f^6}\right).
\end{multline}
We see that the three Goldstones have masses
\begin{equation}
m_1^2 = \beta - \alpha/2\;,\;\;m_2^2=m_3^2 = -\alpha/2,
\end{equation}
and we have, among others, the quartic interactions shown in Fig.~\ref{quartics}.

We can remove another of the Goldstone fields by gauging a further generator of $SO(2)$. For instance, if we gauge
\begin{equation}
T_g^2 = \frac{i}{\sqrt 2} \begin{pmatrix} 0 & 0 & -1 & 0 & 0 \\ 0 & 0 & 0 & 0 & 0 \\ 1 & 0 & 0 & 0 & 0 \\ 0 & 0 & 0 & 0 & 0 \\ 0 & 0 & 0 & 0 & 0 \end{pmatrix},
\end{equation}
then the potential will be exactly as in \eqref{hybridpotential}, with $\phi_3$ set to zero. We must also replace $\beta \rightarrow 2\beta$, since the potential now receives contributions from two gauge bosons.

We note further that if instead we gauged the generator
\begin{equation}
T_g^2 = \frac{i}{\sqrt 2} \begin{pmatrix} 0 & 0 & 0 & 0 & 0 \\ 0 & 0 & -1 & 0 & 0 \\ 0 & 1 & 0 & 0 & 0 \\ 0 & 0 & 0 & 0 & 0 \\ 0 & 0 & 0 & 0 & 0 \end{pmatrix},
\end{equation}
then we obtain
\begin{equation}
V(\phi) = \alpha \cos(\phi/f) + \beta \left(\frac{\phi_1^2+\phi_2^2}{\phi^2} \right) \sin^2(\phi/f) = \alpha \cos(\phi/f) + \beta \sin^2(\phi/f),
\end{equation}
which is symmetric in $\phi_1$ and $\phi_2$.

 \providecommand{\href}[2]{#2}\begingroup\raggedright\endgroup

\end{document}